# The Computing Research Repository:
# Promoting the Rapid Dissemination and Archiving of Computer Science Research

*Joseph Y. Halpern, Carl Lagoze*
*Computer Science Department, Cornell University*
*{halpern,lagoze}@cs.cornell.edu*
http://www.cs.cornell.edu/home/{halpern,lagoze}

## Abstract

We describe the Computing Research Repository (CoRR), a new electronic archive for rapid dissemination and archiving of computer science research results. CoRR was initiated in September 1998 through the cooperation of ACM, LANL (Los Alamos National Laboratory) e-Print archive, and NCSTRL (Networked Computer Science Technical Research Library). Through its implementation of the Dienst protocol, CoRR combines the open and extensible architecture of NCSTRL with the reliable access and well-established management practices of the LANL XXX e-Print repository. This architecture will allow integration with other e-Print archives and provides a foundation for a future broad-based scholarly digital library. We describe the decisions that were made in creating CoRR, the architecture of the CoRR/NCSTRL interoperation, and issues that have arisen during the operation of CoRR.

## 1 Introduction

Computing research relies heavily on the rapid dissemination of results. As a result, the formal process of journal publication has been augmented by other, more rapid, dissemination methods. Originally these involved printed documents, such as technical reports and conference papers. With the advent of the Internet, researchers developed and began to use a variety of electronic means for rapid dissemination. Individual and organizational ftp and web sites made it possible to provide cheap and instantaneous global access to research results.

Immediate access to research results is of tremendous benefit to the research community. However, in many cases it has come at the cost of other features that typically accompany the journal publication process, such as reliable access and high-quality cataloguing, indexing, and archiving. Individual web sites and ftp repositories do not share these attributes.

Cataloguing is essentially non-existent in these repositories. Discovery of their contents entails either visiting the individual sites or using one of the general Internet search engines, which, as described in [6], are not adequate for a number of reasons including lack of domain specificity and failure to index documents in non-textual formats (e.g., PDF). Furthermore, due to *ad hoc* management practices, there is little assurance that the contents of these repositories will persist for any length of time.

A number of efforts over the past several years have addressed these issues with two very different approaches. E-print repositories and organization-specific digital libraries have adopted a *centralized* approach, insuring reliability and longevity through stable funding and good management practices. Architectures to federate separate repositories reflect a *decentralized* approach that permits uniform discovery and access through shared standards and protocols. The respective attributes of each of these approaches suggest that combination of the two -- well-established and well-managed repositories with open interfaces that permit federation -- is the preferred configuration.

This happened in September 1998. Through a partnership of the Association for Computing Machinery[1], the Los Alamos National Laboratory e-Print archive[2] (XXX), and the Networked Computer Science Technical Reference Library[3] (NCSTRL) an online Computing Research Repository[4] (CoRR) was established. CoRR provides a repository in which all members of the community can independently submit papers. CoRR also offers a user interface through which other researchers can browse and search papers currently in the Repository, and subscribe to get notification of new submissions.

Because of its support of the Dienst protocol, CoRR is also integrated into NCSTRL and its contents can be searched, browsed, and subscribed to like any NCSTRL repository. The significance of this open protocol interface to CoRR extends beyond the integration of CoRR documents with the over 27,000 documents currently in NCSTRL. More important is the fact that it lays the foundation for the integration of other repositories both within computer science and across other scholarly fields.

In the remainder of this paper, we provide some background, describe the history of CoRR, its technical foundations, experience with it so far, issues that have arisen as it has developed, and our vision of its future.

## 2 Building on technologies for electronic publishing

There has been a great deal of discussion in the literature of the benefits of electronic publication (and, more generally, electronic access to papers), and the extent to which it can improve over paper publication, both in terms of increased availability, speed of access, and cost of publication (see, for example, [4, 5, 14]). While there is general agreement that electronic access to computer science research would be a good thing, there is much less agreement on how to achieve it. Previous and existing efforts to provide such access can be classified in a number of categories.

***Electronic equivalents of traditional journals:*** Digital libraries run by established publishers or consortia of publishers that typically provide bibliographic searching and full-text access to legacy and new journal issues. Both the ACM Digital Library[5] and the IEEE Computer Society Digital Library[6] are well-known Computer Science examples of this model, which is prevalent in other disciplines as well (for example, HighWire Press[7], Academic Press' IDEAL[8], and JSTOR[9]). These journal-based digital libraries follow a business model similar to the traditional publication system, whereby university libraries and other institutions subscribe. Members of the institution then have access to the documents in the respective digital library.

***Bibliographic servers:*** Centralized indexes that offer searching over the bibliographic records of a number of sources, including journals, conferences papers, and technical reports, such as Alf-Christian Achilles' Collection of Computer Science Bibliographies[10], Michael Ley's Database and Logic Programming Bibliography[11], and David Jones' Hypertext Bibliography Project[12]. In some cases, the bibliographic records contain a URL pointing to content.

***Smart crawlers:*** Programs that scan selected Web and FTP sites and extract bibliographic information and pointers to content at those sites, such as CORA[13], Hypatia[14], and the Unified Computer Science Technical Report Index[15] (UCSTRI).

***Federating architectures:*** Systems that require cooperation from cooperating partners in a common architectural framework, such as Harvest [1], NCSTRL, and the Wide Area Technical Report Service [12]. This cooperation takes various forms including maintaining bibliographic records and content in fixed formats and locations, conforming to established protocols, or running specialized servers.

***Content collectors:*** Systems that download content from various sites and centrally store and index that content, such as the New Zealand Digital Library[16]. The availability of content in a single location permits sophisticated retrieval, automatic classification, and summarization.

***Stand-alone e-Print repositories:*** Repositories to which authors submit papers, which are then accessible by various methods (e.g., browsing, searching) through user interfaces associated with the repositories. Perhaps the most prominent example of this is the Los Alamos XXX e-Print Archive. XXX started as a repository for high-energy physics eprints in 1991, several years before the introduction of the Web. It pioneered the concept of an open-access repository for fast publication of scientific research, and has transformed the dissemination of research in several disciplines. XXX has been quite successful. In August 1998, its coverage included not only most of physics, but also nonlinear sciences, mathematics, and computation and language. It has over 75,000 eprints, is growing at the rate of about 25,000/year, handles over 70,000 transactions/day, has over 35,000 users, and is mirrored in 16 countries. Thanks to funding from the Department of Energy and the National Science Foundation, it also has a full-time staff.

Each of these approaches has achieved some success in providing researchers with easy access to content. At present, however, none of these approaches is dominant in computer science. The resulting fragmentation often is burdensome to the average researcher who is interested in finding, being notified of, and having access to breaking results in his or her field. Furthermore, each of the approaches has had mixed success in addressing a number of issues that are important to consider in the evaluation of a digital library.

***Archiving:*** While it is true that, in rapidly changes fields such as computer science, the most recent research results attract the most interest, the importance of longevity cannot be ignored. Seminal papers attract a considerable amount of interest and, in some cases, the long-term importance of a paper is not obvious until several years after original publication.

***Format Transition:*** An issue related to archiving is how to address the evolution and extinction of digital formats. Competition and fragmentation of the software market has made it almost oxymoronic to talk about a *legacy format*. There has been attention in the literature on mechanisms to address this problem [17], but at present vigilance and manual translation are the best insurance.

***Reliability:*** Availability and consistency of digital content can be compromised by server failure, network partitioning or delays, and questionable or non-existent collection-management standards. Although computer system and network reliability is the subject of considerable research attention, the current reliability of the Internet, WWW, and component servers is often frustrating to the user.

***Extensibility:*** Individuals use information in highly idiosyncratic ways. This is most obvious from the point of view of user interfaces, where even the best user interface in English may be completely useless to a user in Japan. It is also true with other less "visible" aspects of information use, such as organizational schemes and collection definition. The definition of "computer science" from the point of a faculty member at a major university differs substantially from that of one at a community college. This need for flexibility and recognition of individual information needs has influenced much of the interoperability research within the digital library community [11, 16, 18].

Broadly speaking, the ability to address these issues reflects the tension between centralized and decentralized technologies. In general, effective archiving, attention to format transition requirements, and good reliability is best achieved in relatively centralized environments where there can be a relatively high degree of control and oversight by experienced staff. The Los Alamos archive, with a stable source of funding and a committed and experienced staff, is able to do this. It is much more difficult with highly decentralized technologies, such as smart crawlers. A distributed approach such as NCSTRL, is able to do this to some extent, but as described in [3], the system suffers due to varying levels of follow-through on the part of its members. While some departments are quite scrupulous in the maintenance and management of their servers, others pay little attention to this task. As a result, the perceived performance of the entire system suffers.

On the other hand, when it comes to extensibility, decentralized systems typically do much better. Self-contained centralized systems typically do not provide mechanisms for interoperation. In software engineering terms, these systems do not generally offer an API through which the functional components of the system can be accessed and manipulated by external programs and agents. The lack of an API makes it difficult to replace components of these systems or provide value-added services. Decentralized systems, especially those that involve the common usage of a suite of protocols (such as Dienst), have built in extensibility because the interoperation of the system components is based on the exposure of low-level functionality. Components can therefore be replaced, redefined, and enhanced in a modular fashion.

We hesitate to argue that there is a single *correct* technology for scholarly electronic publishing. Like many digital library issues, technical solutions must co-exist within organizational contexts. For example, the arguments for a highly centralized approach should recognize the reality that individual organizations may wish to maintain physical control over their document repositories. Arguments for a decentralized approach must acknowledge the frequent lack of management standards at individual repositories. We suggest that a desirable approach for a reliable and extensible scholarly publishing architecture is a middle ground involving the federation of a set of well-managed repositories. These considerations greatly influenced the development of CoRR, as described in the next section.

# 3 Setting up CoRR: issues and decisions

ACM, was (and continues to be) interested in experimenting with different approaches to disseminating research. In May 1997, a committee[17] was formed under the auspices of the ACM Publications Board to consider one such approach: setting up an online repository for computing research. Initially, the main focus of the committee's discussions revolved around the design of the architecture. Three main options emerged.

The first option was to become part of the Los Alamos XXX repository. As a base for a computing repository, XXX had many attractive features, perhaps the most important of which was that it clearly worked and worked well. However, the ACM committee decided against this option primarily because XXX did not provide an open interface and was therefore not amenable to extension and enhancement.

The second option considered was to become a node in NCSTRL by running a Dienst server. The most important features of NCSTRL from the committee's point of view were that it was explicitly designed with an open interface and it was a computer science effort. On the other hand, Dienst servers are designed to manage documents within an institution or organization and, for example, do not permit open submissions or moderation.

The third option was to build a new system from scratch. This had the obvious advantage that we could design our own system, which hopefully would have exactly the attributes we required, but had the equally obvious disadvantage that it would take time, money, and expertise.

The committee settled on a hybrid approach that combined the best features of XXX and NCSTRL, using the well-tested XXX software for submission, notification, and searching, while still taking advantage of the open NCSTRL architecture. From the point of view of the NCSTRL interface, XXX is now just a node on NCSTRL. (The technical details of the architecture design are discussed in Section 4.)

With the major decision out of the way, there were still a number of other important decisions that had to be made regarding how CoRR would operate:

***How should CoRR be organized?*** The physics and mathematics archives at XXX are organized into a relatively small number of subject areas -- 38 in the case of physics and 31 in the case of mathematics. These subject areas play a number of roles. From the perspective of document submissions, they form the basis for moderation; that is, for each subject area, there is a moderator who checks submitted papers for relatedness to the subject (although not quality or novelty). At the user interface level, they are used as aids for searching, browsing, and subscribing.

The committee had to decide how to partition the computing field into subject areas. One choice was to use the ACM classification system[18]; the alternative was to create a new classification system. The ACM classification scheme has the advantage of being a relatively stable scheme that covers all research in computing, which has been carefully crafted over the years. Unfortunately, it does not seem to map too well to the current major areas in academic computer science. In particular, it seemed to be difficult to find moderators for subject areas that corresponded to major topics under the ACM classification system.

In the end, the committee chose to use both approaches: authors are asked to classify papers both by using a subject area from a list of 33 subject areas[19] and by choosing a primary classification from among the roughly 100 third-level headings in the 1998 ACM Computing Classification scheme. While the subject areas are not mutually exclusive, nor do they (yet) provide complete coverage of the field, they seem to better reflect the current active areas of research in CS. Each subject area has a moderator. While documents are partitioned by subject area, readers can search and get notified of new papers both by ACM classification and subject area.

Interestingly, the committee that formed the mathematics archive at XXX chose not to use the AMS classification scheme at all, instead opting for their own subject classification[20]. Their reasons for designing a separate subject classification were quite similar to ours.

***What about copyright?*** Publishers typically require authors to transfer copyright when they publish a paper, since it gives them more freedom of action and more control of the disposition of the paper. The committee decided not to require any transfer of copyright or publication rights. Authors will continue to retain copyright when they submit (although they may have to transfer rights if they wish to publish in certain journals).

***How long should papers stay on CoRR?*** CoRR is intended to be archival; the expectation is that papers submitted will stay there permanently. This does not prevent authors from updating their papers. Updated versions of a paper can be posted at any time, but earlier versions of the paper remain on the repository as well. All versions are timestamped, to avoid confusion. The most recent version of the paper is the one accessed by default, but there will be pointers to the earlier versions. This prevents a situation where, for example, author A improves on the results in an early version of B's paper, but finds that these improvements seem foolish when the only version of B's paper that is available has better results.

***What submission formats should be accepted?*** Currently, authors can submit documents to CoRR using TeX/LaTeX/AMSTeX, HTML+GIF, PDF, or Postscript. However, if TeX (or one of its variants) is available, following the policy at XXX, authors are required to submit the source. If an author has generated Postscript or PDF from some variant of TeX, it is rejected in favor of the TeX source (which is then used to generate Postscript for a number of platforms and PDF). While there are good reasons[21] for this policy -- for example, the source file seems much better for archival purposes -- it has generated perhaps the most controversy (see Section 5).

# 4 Building interoperability between XXX and NCSTRL

As we said, CoRR was intended to combine the best features of XXX and NCSTRL. In this section, we describe how this was done.

NCSTRL is a federated collection consisting of a set of independent repositories and digital library servers that interoperate using the Dienst protocol. We briefly describe the Dienst protocol here; more details can be found in [2, 8, 9].

The Dienst architecture consists of a set of service definitions and a document model that are exposed through the service requests defined by the Dienst protocol. The core services defined by Dienst can be divided into two groups:

1. Self-contained services, that do not communicate with or depend on other services. These include a *Repository Service* that stores digital documents, an *Index Service* that accepts queries and returns lists of documents identifiers matching those queries, and an *Info Service* that returns information about the state of a server hosting services.

2. Services that logically combine or use other services, interacting with them via the Dienst protocol. These include a *Collection Service* that provides information on how a set of services (and servers) interact to form one or more collections, a *Registry Service* that stores information about (human) users of services of one or more collections, and a *User Interface* service that mediates human access to all the other services.

The Dienst definition of services and corresponding service requests enables flexible and extensible interoperability. For example a complete distributed digital library can be created from a set of sites, each running one or more Dienst services, with at least one repository service and one index service among the set of sites. While users can interact with this digital library through a Dienst user interface service, other clients and agents can communicate with any of the component services using raw Dienst protocol requests. Furthermore, a completely new user interface to this digital library and its components can be constructed that presents a different view of the library contents (for example, in a different language). Finally, new value-added services (e.g. summarization, payment services) can be constructed that interact with the core services.

There are a number of ways that interoperability could have been attained between the XXX archive and NCSTRL. For example, it would have been possible to use what Paepcke *et.al.* [15] call *external mediation*. In this scenario, there would be proxy software at the Dienst NCSTRL side that translated Dienst requests to XXX HTTP requests and parsed the HTML responses from XXX to create Dienst protocol responses. While this would have worked and required no intervention by LANL staff, it was deemed undesirable, for two main reasons.

*It is inherently unstable.* Lacking agreement on a formal protocol for interoperability renders it vulnerable to changes. Problems with such proxies most frequently occur at the response rather than the request end of the communication. For example, a small change in an HTML response by XXX (such as improving the appearance of a browser page) might break the parsing mechanism that translates to a corresponding Dienst protocol response.

*It does not scale or provide a foundation for future interoperability with other archives.*

Such proxies are an *n by n* approach to interoperability, whereby each site interoperates with another through idiosyncratic methods.

A more preferable model of interoperation, especially given the similarity of operational semantics between NCSTRL and XXX, is using what Paepcke *et. al.* [15] call *strong standards*, in this case the Dienst protocol version 5.0. Staff at XXX agreed to support the following Dienst service requests:

### Repository Service Requests

- *List-Partitions* - return the partitions (subject areas) by which the repository is divided (e.g., artificial intelligence, databases etc.).

- *Formats* - return the formats available for a particular document.

- *List-Versions* - return the versions available for a particular document.

- *Disseminate* - return a specified version of a document in a specified format.

### Index Service Requests

- *List-Contents* - return bibliographic records in RFC1807 format [10] (this is the attribute/value metadata format used by NCSTRL).

The Dienst index service *SearchBoolean* request is notably missing from this list. This is because the XXX system does not at this time support metadata field-based boolean searching (e.g., *author contains Hartmanis and title contains computational complexity*).

Implementation of these requests permits the following level of interoperability. The XXX repository is, from the point of view of all NCSTRL sites, supporting protocol version 5.0, just like any other Dienst repository. Using the repository service messages listed above, NCSTRL user interfaces are able to access information about and retrieve documents from the XXX repository. Without support for the Dienst *SearchBoolean* request at XXX (for reasons noted above), searching the contents of the XXX archive must be done at another NCSTRL search engine. This flexible routing of search requests, however, fits naturally into the Dienst collection service architecture [8].

Figure 1 illustrates this interoperability architecture. The solid lines connecting the user interface to the index servers and XXX repository are Dienst protocol interactions during a user's search and retrieval of a document. The broken black line indicates the periodic download of bibliographic records by a Dienst index server from the XXX repository. Finally, the solid lines connecting the WWW browser to other components indicate HTTP protocol requests to both the Dienst user interface server (for Dienst type search and retrieval of documents) and to the XXX repository (for document submission to that repository and XXX internal search and retrieval).

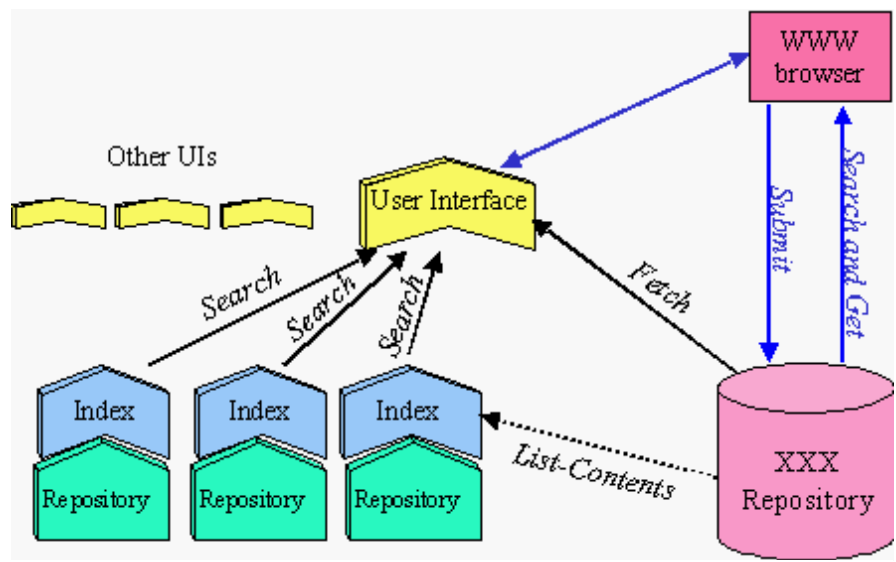

**Figure 1 - Interoperability between XXX and NCSTRL**

# 5 Issues Raised through Experience

CoRR has been in operation since Sept. 15, 1998. After an initial flurry of over 100 submissions, the recent rate has been slightly over one submission per day. Currently there are about 1200 papers in CoRR, with about 900 of them coming from a previously-existing archive at XXX on Computation and Language, which has now been folded into CoRR. These are combined through the NCSTRL interface with over 27000 other papers. It is too early to tell whether CoRR will really catch on. However, a number of groups have tentative plans to migrate archives to CoRR and to use CoRR as a repository for journal and conference activities (including DL '99) so the prognosis is good.

Even in the few months of operation, a number of issues have arisen.

***Insistence on source files:*** There have been bitter complaints from users about the insistence on TeX source. There have been two orthogonal reasons for these complaints.

The first involves ease of submission. There is no question that it much easier to submit a single Postscript file than it is to submit a TeX file and a number of auxiliary files (such as figures in postscript, a bibliography file, and macro files). While we hope that scripts will be developed that will automatically collect all the relevant files, sparing authors the burden of doing so, we continue to insist that the source file(s) be submitted.

The second reason is that authors are concerned with the fact that source files are available for download (although, in fact, they are rarely downloaded -- readers prefer Postscript or PDF). Authors are concerned that the availability of the source will make plagiarism easier (note that it is certainly possible to plagiarize even from PostScript, given the availability of postscript to ASCII converters) and also give readers access to comments that were intended to be private. To deal with these concerns, the CoRR advisory committee has agreed to give authors the option of making their source unavailable to readers, although the repository will still retain the source and so as to be able to use it to convert to new formats as they arise.

It is interesting to note that the physics community (which presumably has much the same concerns as the computing community) has been submitting source files for years. This suggests that there may be cultural differences between the communities; alternatively, perhaps in time the computing community will also become more comfortable with submitting source files.

***Departmental Technical Reports:*** Some departments are debating whether to maintain technical reports in their own NCSTRL-accessible repository, or to encourage members to submit to CoRR and maintain a list of pointers to these papers at their own departmental site. From the perspective of users searching and accessing CoRR through NCSTRL this choice is largely transparent.

This issue is related to our earlier discussion of centralized vs. decentralized technologies. Departments that do not have (or are unwilling to commit) the resources to maintain and manage their servers are better off using the services of CoRR for document storage. XXX has the resources to ensure that their servers are almost always available (and, in any case, there are many mirror sites). There are definite economies of scale to be gained by having many papers at one site. On the other hand, by maintaining their repositories, departments will have greater control over the storage of and presentation of their documents (especially if they choose to participate in NCSTRL by running a Dienst server, rather than simply maintaining an ftp repository in the fashion of NCSTRL LITE[22]).

Clearly this is an issue that has to be decided on a department-by-department basis. Whatever decision individual departments make, it will almost certainly be the case that there will be some individuals will submit papers to CoRR and to their departmental server, which means that an NCSTRL search will locate the same paper (or variants of the same paper) twice. This is an issue that will have to be dealt with through some mechanism such as common unique identifiers for the multiple copies.

***Copyright and journal publication***: There are fields (such as medicine and chemistry) for which publishers will not publish papers that have appeared on the web (even on an author's personal web site). This has not been the case in computer science, and is unlikely to become so. Researchers have come to expect that they will be able to make their papers available rapidly at online sites, such as CoRR, while still submitting their papers to conventional journals. However, publishers may insist, as part of their copyright policy, that a paper be withdrawn as a precondition to journal publication. In this case, at the author's request, the paper will indeed be withdrawn. However, we are hoping that this will not happen. ACM, in particular, has agreed that, at least for the next two years, authors can leave a version of the paper on CoRR that is essentially identical to one that appears in an ACM journal. (It will also be possible to link to the definitive version on the ACM Digital Library.) This is an experiment -- ACM is examining the impact that this will have on journal sales. We hope that other publishers will consider similar experiments.

***Funding***: Currently, CoRR is riding on the coattails of NSF and DARPA funding provided to XXX and NCSTRL, and this should suffice for the foreseeable future. The long-run funding situation is not yet clear. Clearly, when a resource becomes as important to a community as the XXX archives are to physicists (and we hope that CoRR will be to computer scientists), that community will collectively work to ensure funding. However, the economic models for electronic scholarly publishing (and for the Internet as a whole) are still the subject of considerable investigation [13, 19]. It is not clear how the funding issue will be resolved in the long run.

# 6   Where we go from here

This is a period of change in scholarly publishing and nobody can predict the changes that will happen over the next few years. The impact of CoRR and similar efforts on conventional journals is, no doubt, a question that many journal publishers are asking. There are a number of possibilities. In one scenario eprint repositories such as CoRR could co-exist with the conventional journal model (recognizing that that model will undoubtedly move to electronic dissemination). For example, XXX has been providing eprint archives in physics since 1991 without apparent impact on conventional journals. In another scenario, efforts such as CoRR could provide the foundation for a new and enhanced role for conventional publishers. Publishers could provide value-added services to authors and readers -- such as summarization services, advanced searching tools, awareness, and filtering services -- that build on the content in CoRR-like repositories. Clearly, if efforts such as CoRR develop into the primary vehicle for dissemination of research results, they could significantly change the business model for scholarly publication.

We feel strongly that the CoRR model of high-integrity repositories that provide interfaces for federation with other repositories will play a critical role in the emergence of global digital libraries for scholarly information. This model has the advantage of providing the stable administrative environment for sufficient management and evolution of the content maintained in the repositories, while at the same time allowing the expression of specific community and organizational interests.

There are a number of efforts in other disciplines to develop on-line scholarly publishing archives. We are optimistic that the effort and architecture described here will serve as a model for broader interoperability among these archives. The federation of these archives could be an important step towards the creation of the National Science, Mathematics, Engineering, and Technology Digital Library (NSDL) described in [20]. Organizing such a large multi-disciplinary collection so as to maintain usefulness to the user is nontrivial. We believe it will have to be done in such a way that an individual user will be able to decide how to view (that is, browse and search) the collection, and what portions of it to view. Without such functionality, the collection has the danger of moving in the direction of the World Wide Web where it is all too easy for valuable information to be lost in the noise.

The development of a growing and multi-faceted research collection provides a unique testbed for future digital library research. The participation of NCSTRL (and by inclusion CoRR) in the DARPA-funded D-Lib Test Suite program[23] provides the framework for this type of research. In this vein, we suggest the following example research and implementation areas.

***Improved and specialized user interfaces:*** The NCSTRL and XXX interfaces to the computer science collection are rudimentary and could certainly be greatly improved, both in terms of functionality and ease of use. The open architecture described here facilitates experimentation with new user interfaces that conform to different community needs. There are tentative plans for building a new interface gateway at the ACM web site.

***Overlay journals:*** A bare-bones journal could be built that would simply be a collection of pointers to documents in CoRR and other federated repositories. The journal would have an editorial board just as journals do now. Rather than (or in addition to) coming out in print, once a paper was accepted, the final version would be deposited in CoRR, and there would be a pointer to it from the journal's web site. There currently is one such overlay journal in physics - Advances in Theoretical and Mathematical Physics[24]; discussions are being held to set up such journals for computer science as well.

***Comments:*** A comment facility could be added to allow readers to add comments to papers.

Such comment facilities have met with mixed success in the past. For example, there has been little usage of the comment facility provided by the Journal of Artificial Intelligence Research[25]; the comment facility at the Electronic Transactions on Artificial Intelligence[26] has seen more usage.

***Specialized collection definition:*** The availability in federated repositories of a large number of research papers from a variety of disciplines, as envisioned here, makes possible the definition of specialized sub-collections – for example, a collection tailored to a specific university course or curriculum. As described in [7], these collections might be defined by criteria that cross standard disciplinary boundaries and include documents in distributed archives.

***Document cross-linking:*** Mechanisms to both assist in linking related documents and create those linkages automatically could enhance the ability of researchers to discovery related research.

***Current awareness services:*** Both the NCSTRL and XXX user interfaces include subscription facilities that alert researchers of new additions to the collection. These current facilities are useful but crude. One approach ripe for experimentation in this environment could be collection of user profiles (both manually and automatically) and the use of those profiles in current awareness services.

We are optimistic that, given CoRR's open interface, these enhancements and others will be investigated and implemented by the efforts of many researchers in the community.

# Acknowledgements

We would like to acknowledge the contributions of the CoRR committee and the ACM publications board, especially Bill Arms and Peter Denning, for their role in setting up and supporting CoRR. Work on NCSTRL is supported by the Defense Advanced Research Project Agency under Grant No. MDA 972-96-1-006 and N66001-98-1-8908 with the Corporation for National Research Initiatives. This paper does not necessarily represent the views of CNRI or DARPA. Some material in this article appeared in preliminary form in J. Y. Halpern, "A Computing Research Repository", *D-Lib Magazine* (November 1998).

## Hyperlinks

[1] http://www.acm.org

[2] http://xxx.lanl.gov/

[3] http://www.ncstrl.org

[4] http://xxx.lanl.gov/archive/cs/intro.html

[5] http://www.acm.org/dl

[6] http://www.computer.org/epub

[7] http://www.highwire.org

[8] http://www.idealibrary.com

[9] http://www.jstor.org

[10] http://liinwww.ira.uka.de/bibliography/index.html

[11] http://dblp.uni-trier.de

[12] http://theory.lcs.mit.edu/~dmjones/hbp

[13] http://www.cora.justresearch.com

[14] http://hypatia.dcs.qmw.ac.uk

[15] http://www.cs.indiana.edu/cstr/search

[16] http://www.nzdl.org/cgi-bin/gw

[17] http://www.acm.org/corr

[18] http://www.acm.org/class/1998/overview.html

[19] http://xxx.lanl.gov/archive/cs/subj.html

[20] http://xxx.lanl.gov/new/math.html

[21] http://xxx.lanl.gov/help/faq/whytex

[22] http://lite.ncstrl.org:3803/Dienst/htdocs/Lite/ncstrl-lite.html

[23] http://www.dlib.org/test-suite

[24] http://www.intlpress.com/journals/ATMP

[25] http://www.cs.washington.edu/research/jair/home.html

[26] http://www.ida.liu.se/ext/etai